# Effets d'une suppléance perceptive visuelle, auditive et tactile sur le contrôle des pressions fessières en position assise

**Olivier Chenu\*,\*\* — Rémy Cuisinier\* — Nicolas Pinsault\* — Jacques Demongeot\* — Yohan Payan\* — Nicolas Vuillerme\***

*\* Laboratoire TIMC-IMAG*
*Faculté de Médecine*
*38706 La Tronche cédex*

*{prenom.nom}@imag.fr*

*\*\* IDS SA*
*56, Quai Jules ChagotBP 107*
*71304 MONTCEAU LES MINES CEDEX*

RÉSUMÉ. *Cet article présente une étude de différentes modalités informatives d'un dispositif de suppléance perceptive destiné à réduire les surpressions fessières prolongées. Les modalités visuelles, auditives et tactiles sont analysées et comparées à une situation sans retour sensoriel. En conclusion, si les différentes modalités ont un effet positif et comparable, elles sont diversement jugées en terme de confort et de perturbation vis-à-vis d'autres activités.*

ABSTRACT. *This article presents a study on different informative modalities of a percecptual supplementation device aiming at reducing overpressure at the buttock area. Visual, audio and tactile modalities are analysed and compared with a non-biofeedback session. In conclusion, modalities have a positive and equal effect, but they are not equally judged in term of comfort and disturbance with some other activities.*

MOTS-CLÉS : *substitution sensorielle, suppléance perceptive, tactile, electro-tactile, prévention escarres, surpressions.*

## 1. Introduction

### 1.1. *Les escarres*

Une escarre est une plaie plus ou moins profonde provoquée par une compression des tissus (éventuellement associée à une friction ou un cisaillement) [http://www.epuap.org/gltreatment.html] ; en effet une forte pression prolongée prive les tissus internes d'irrigation sanguine et ceux-ci meurent (figure 1).

Le traitement des escarres est coûteux en termes psychologique et sociétal pour la victime puisqu'il peut durer plusieurs mois (Garber et al, 2003) et impliquer un alitement total. Par conséquent, le coût financier est lui aussi très lourd (Barrois, 1998).

Les principales victimes des escarres sont les patients en cours d'hospitalisation et les personnes blessées médullaires. On estime que plus de 80% de ces dernières développent en effet au moins une escarre au cours de leur vie (Byrne et al, 1996) (Salzberg et al, 1996). Privées de leurs retours sensoriels au niveau des segments inférieurs du corps du fait de la blessure médullaire, ces personnes restent assises sur leurs fauteuils roulants la plupart du temps et ne perçoivent plus les messages d'alerte voire de douleur qui devraient leur permettre d'éviter la formation d'escarres. Ces informations sont en permanence perçues par les personnes valides sous forme de « fourmillements » qui les obligent à se mobiliser ou à soulager leur assise régulièrement.

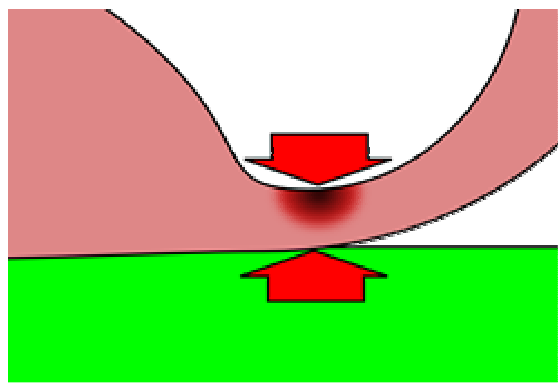

**Figure 1.** Une escarre est une plaie profonde produite généralement par la compression des tissus contre les os.

Différentes approches ont été proposées pour prévenir la formation d'escarres (Reddy et al, 2006), toutes sont basées sur des systèmes passifs de redistribution des pressions (coussins) ou sur des soins infirmiers (repositionnement des patients, optimisation de l'état nutritionnel et hydratation de la peau dans les zones à risque), ces derniers étant surtout dédiés aux cas des hospitalisations. Au jour d'aujourd'hui, aucune méthode de prévention ne s'est avérée réellement efficace. D'ailleurs Chenu et al (Chen et al, 2005) observe même une augmentation de la prévalence des escarres chez le blessé médullaire dans une étude de 1994–2002 par rapport à une précédente de 1984–1993.

Le laboratoire TIMC-IMAG a décidé de se pencher vers un principe qui a fait ses preuves dans de nombreux domaines : la substitution sensorielle ou l'art de suppléer une modalité sensorielle défaillante par une autre modalité sensorielle.

**1.2.** *La substitution sensorielle*

Le principe de substitution sensorielle a été introduit et étudié par Paul Bach-Y-Rita qui a montré que des stimuli caractéristiques d'une modalité sensorielle peuvent être transformés en des stimuli d'une autre modalité sensorielle tout en restant correctement interprétés (Bach-y-Rita et al, 1969). Dans le cas de la substitution visuelle que Bach-Y-Rita a largement étudiée, des images numériques acquises par une caméra attachée au front de sujets aveugles étaient échantillonnées et affichées sur une matrice tactile placée en divers lieux du corps (main, dos, ventre, front et langue). Deux types de dispositifs ont ainsi été testés (un dispositif vibro-tactile et l'autre électro-tactile) et ont montré leur efficacité dans le cadre de la substitution visuelle. Le dispositif électro-tactile ayant certains avantages en termes de consommation électrique et d'encombrement, l'équipe de Bach-Y-Rita a finalement convergé vers un dispositif d'électro-stimulation linguale : le « Tongue Display Unit » (TDU) (Bach-y-Rita et al, 1998). La langue possède en effet nombre d'avantages en terme de canal sensoriel: c'est un organe fortement innervé et très sensible.

**1.3.** *La substitution sensorielle appliquée à la prévention des escarres*

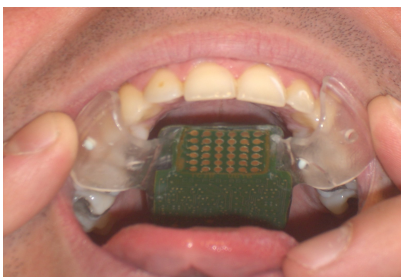

**Figure 2.** Le TDU embarqué

Forts des résultats encourageants de Bach-Y-Rita, le laboratoire TIMC-IMAG a développé son propre TDU embarqué. Inclus dans un palais orthodontique, celui-ci est alimenté par une pile-bouton interne et communique avec un ordinateur distant grâce à un récepteur RF (figure 2).

Cette étude vise à évaluer l'utilisation de ce dispositif embarqué pour limiter les surpressions prolongées mesurées au niveau de l'assise. L'objectif final est la mise au point d'un dispositif de prévention des escarres en direction des paraplégiques, c'est-à-dire des personnes qui ne perçoivent plus les surpressions fessières mais qui conservent la mobilité du buste. Le principe du dispositif consiste à mesurer les surpressions, à estimer le changement postural du buste qui réduira au mieux ces surpressions puis à renvoyer ce message de changement postural à la personne. Nous avons décidé de placer les sujets étudiés dans une situation potentiellement « à risque », c'est-à-dire une activité qui demande une charge attentionnelle importante, en l'occurrence le visionnage d'un film. L'objectif de notre étude est de comparer ce dispositif électro-tactile à des dispositifs de renvois d'informations plus usuels (en l'occurrence visuels et sonores) (figure 3) en termes de performance et d'acceptation.

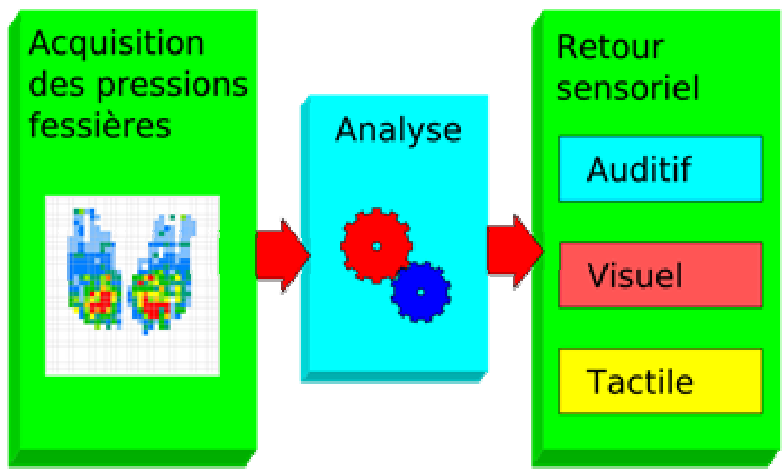

**Figure 3.** Le principe de la substitution sensorielle appliquée à la prévention des escarres

## 2. Méthode et Matériel

### 2.1. *Fonctionnement du dispositif*

#### 2.1.1. *Acquisition des pressions*

Fournie par la société Vista Medical, une nappe de pression flexible installée sur un siège permet d'acquérir les données de pression à l'interface siège/peau. Cette nappe consiste en une matrice de 1024 capteurs (32x32), chacun pouvant mesurer des pressions entre 0 et 200 mmHg sur une surface d'environ 2cm².

Les ergothérapeutes des centres de rééducation utilisent de telles nappes pour évaluer l'efficacité de certains coussins, ainsi que pour informer visuellement les patients paraplégiques des dangerosités de certaines postures. Malheureusement, privé de cet outil dans sa vie quotidienne, le paraplégique reprendra assez vite de mauvais comportements.

#### 2.1.2. *Détection des surpressions et estimation d'un changement postural*

Le but de cette étude est de montrer que des dispositifs de substitution sensorielle peuvent permettre de faire disparaître des surpressions apparues au long d'une période. Soient P_t la matrice de pression à l'instant t et P_M la matrice de pression moyenne acquise pendant une certaine durée. Soit HP_M la matrice égale à la nappe P_M en tout point où la pression est supérieure à 100 mmHg et valant 0 partout ailleurs. La matrice HP_M représente les surpressions que l'algorithme doit essayer de faire disparaître.

Pour ceci, l'algorithme a deux tâches : estimer la posture actuelle du sujet et la posture la plus susceptible de supprimer les surpressions emmagasinées jusqu'alors. Pour ceci, trois matrices auront été pré-enregistrées avant l'expérimentation. Nous les appellerons PR_C, PR_G et PR_D et sont acquises alors que nous aurons demandé explicitement au sujet de se mettre respectivement dans une posture Centrale, penché à Gauche et penché à Droite.

Nous définissons la distance de Manhattan d de la façon suivante :
$$d(A, B) = \sum_{capteurs\_c} |A(c) - B(c)|$$

Où A(c) (resp. B(c)) est la valeur du capteur c de la carte de pression A (resp. B).

Ainsi la posture X (X valant C, G ou D) dont la matrice correspondante PR_X a une distance d(P_t, PR_X) minimale est considérée comme la posture dans laquelle se trouve le sujet à l'instant t.

Enfin, pour déterminer la posture la plus apte à réduire les surpressions HP_M, nous introduisons la distance pondérée wd suivante :
$$wd(A, B) = \sum_{capteurs\_c} t(A(c)) \cdot (A(c) - B(c))$$

Où t(A(c)) = max(A(c)-T, 0) avec T valant 100 mmHg, une plus grande importance étant considérée pour les hautes pressions.

La posture cible Y (Y valant C, G ou D) est celle pour laquelle la matrice correspondante PR_Y a la distance wd(HP_M, PR_Y) maximale.

### 2.1.3. *Diriger le sujet vers la posture cible*

Le message à renvoyer au sujet s'en trouve alors très simplifié. Connaissant la posture d'origine et la posture cible, il suffit donc d'indiquer à l'utilisateur une direction vers laquelle se pencher (Gauche ou Droite). Cette simplicité va se traduire par une simplicité du dispositif de renvoi d'information et va donc nous permettre de tester relativement facilement différents retours sensoriels : la vue, l'ouïe et le tactile. L'information auditive sera concrétisée par un signal sonore dans l'écouteur Gauche (ou Droit) d'un casque stéréophonique, l'information visuelle par une couleur rouge sur les bords Gauche (ou Droit) d'un écran d'ordinateur et l'information électrotactile par l'activation des électrodes du bord Gauche (ou Droit) de la matrice du TDU embarqué.

De plus, parce que le message envoyé représente un signal de « douleur », celui-ci sera opposé au changement de posture désiré ; par exemple, un signal à Droite signifie un message de « douleur » à droite nécessitant un déplacement vers la Gauche.

### **2.2.** *Expérimentation*

Rappelons que cette étude vise l'évaluation de l'efficacité et l'acceptation de trois dispositifs de biofeedback par suppléance perceptive pour réduire les surpressions pendant une tâche supposée à risque car demandeuse d'attention et totalement passive : le visionnage d'un film.

### 2.2.1. *Les sujets*

8 jeunes sujets sains ont participé à cette étude (7 hommes/1 femme, 27 ans +/- 4, 173 cm +/- 7, 69 kgs +/- 5) ainsi qu'un sujet paraplégique et ont donné leur consentement.

### 2.2.2 *Procédure*

Du fait que la sensibilité linguale est très différente d'un individu à l'autre, la première phase de l'expérimentation consistait à calibrer les intensités délivrées par le TDU (entre 0 et 5V) de façon que le sujet perçoive un signal fort mais non douloureux.

Après quoi le sujet s'assoit sur une table sur laquelle est disposée la nappe de pression. L'expérimentateur lui demande alors explicitement de se placer dans trois postures types (Gauche, Droite et Centrale) afin d'enregistrer les matrices de pression PR_G, PR_D et PR_C précédemment citées.

Viennent ensuite les sessions de tests pendant lesquelles le sujet est équipé d'un casque stéréophonique et regarde un film sur un écran d'ordinateur placé devant lui. Chacune de ses sessions dure 7 minutes ; il y en a 4 au total espacées de quelques minutes de repos. Les 4 sessions sont respectivement passées sous les conditions suivantes : suppléance tactile (pendant laquelle le sujet est équipé du TDU), suppléance visuelle, suppléance auditive et aucune suppléance perceptive (condition contrôle). Lors de ces sessions de suppléance, le sujet perçoit un signal toutes les 20 sec +/- 4s et doit se déplacer dans la direction opposée au signal jusqu'à ce que celui-ci disparaisse. En cas d'absence de réaction ou de mauvaise réaction du sujet, le signal s'éteint de lui-même au bout de 10 secondes.

A l'issue de l'expérimentation, un questionnaire est adressé au sujet. Pour chaque modalité sensorielle, il doit indiquer à quel niveau il juge le signal confortable (de 0 signifiant « très confortable » à 10 signifiant « très inconfortable ») ; il doit indiquer aussi dans quelle proportion le signal a perturbé son visionnage du film (0 : pas de perturbation ; 10 : énormes perturbations) et réciproquement si le visionnage du film a pu gêner la perception du signal (0 : pas de gêne ; 10 : énorme gêne). Finalement, il lui est demandé de classer les modalités dans l'ordre de ses préférences.

2.2.3. *Analyse des données*

La première variable analysée est la réussite ou non de la tâche d'atteinte d'une posture cible. En d'autres termes, la posture finale (à la fin de la stimulation) est elle la posture cible visée par l'algorithme ?

De plus cette réussite s'accompagne t'elle réellement d'une baisse des surpressions ? Considérons la zone de risque Arisk qui couvre l'ensemble des capteurs de HP_M dont la valeur est supérieure à 0. Cette zone est celle dans laquelle les capteurs étaient en surpression avant la stimulation sensorielle et dont on se donne comme objectif de réduire les niveaux. Considérons ensuite le volume de surpressions VS comme la somme des valeurs des capteurs dans Arisk. Une seconde variable dépendante en a été tirée : la réduction du volume VS après la stimulation sensorielle par rapport à avant ; nous appellerons celle-ci VVS (pour Variation du Volume de Surpression)

## 3. Résultats

### 3.1. *Efficacité*

La figure 4 montre dans quelle mesure l'ordre sensoriel a été correctement perçu et interprété. En d'autres termes, la posture en fin de stimulation estimée par l'algorithme est-elle bien la posture cible susceptible de diminuer au mieux les surpressions précédemment emmagasinées. Dans 24 cas sur 27 (toutes modalités confondues), toutes les stimulations semblent avoir été perçues puisque la posture finale est bien celle attendue par l'algorithme.

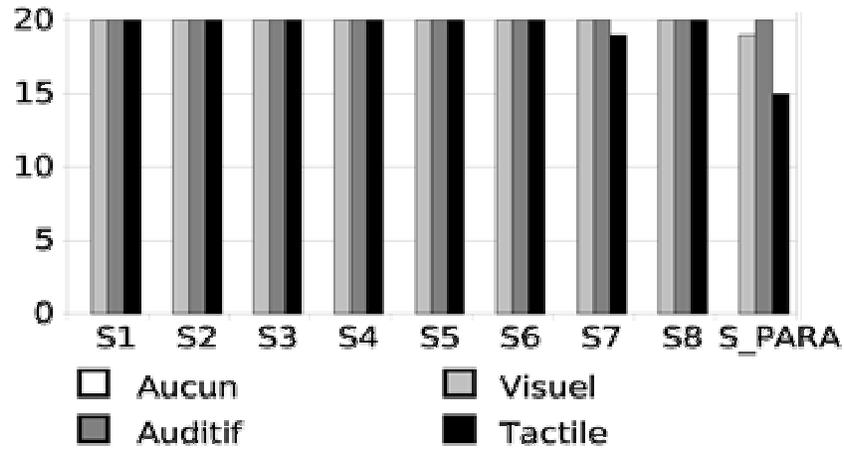

**Figure 4.** Perception et compréhension du signal

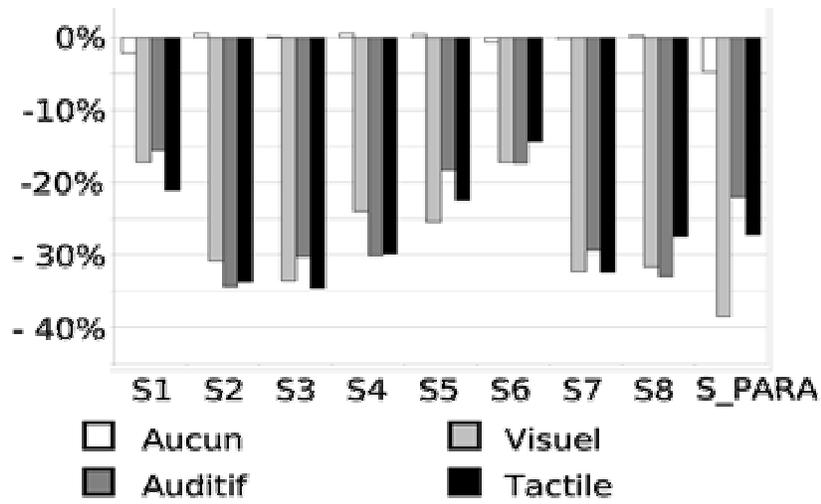

**Figure 5.** Variation du Volume de Surpression en pourcentage

L'effet du biofeedback est ici indiscutable car dans toutes les conditions sans biofeedback, aucune posture cible n'a été atteinte. Dans la pire des conditions (sujet paraplégique équipé du dispositif tactile), on peut lire que 75% des signaux ont été correctement perçus.

La figure 5 montre quand à elle la variable VVS dans chacune des conditions. Ainsi, dans la condition sans biofeedback, la variation est comprise entre -5% et +2% ; dans toutes les autres conditions, cette variation reste strictement négative et est inférieure à -14%

### 3.2. Ergonomie

Le tableau 1 montre les réponses des sujets au questionnaire final sur le confort et la gêne de chaque modalité ainsi que leurs préférences. On peut y lire que l'information visuelle est considérée comme la plus confortable, mais que le visionnage du film semble perturber la perception du signal ; l'information auditive, quant à elle, semble gêner le visionnage du film ; enfin le TDU, s'il a globalement de bonnes critiques vis-à-vis de la gêne avec le film, est encore très mal accepté au niveau confort. Il s'ensuit que les modalités préférées sont, par ordre décroissant la modalité visuelle, auditive et tactile.

| question   |   | S1    | S2    | S3    | S4    | S5    | S6    | S7    | S8    | S PARA | MOYENNE |
|------------|---|-------|-------|-------|-------|-------|-------|-------|-------|--------|---------|
| Confort ?  | A | 2     | 3     | 4     | 5     | 3     | 4     | 4     | 3     | 2,5    | 3,39    |
|            | V | 1     | 2     | 3     | 3     | 1     | 2     | 3     | 0     | 1      | 1,78    |
|            | T | 5     | 7     | 6     | 4     | 6     | 6     | 3     | 5     | 8      | 5,56    |
| SPF ?      | A | 3     | 2     | 6     | 2     | 3     | 4     | 2     | 4     | 2,5    | 3,17    |
|            | V | 2     | 1     | 1     | 0     | 1     | 2     | 2     | 1     | 1      | 1,22    |
|            | T | 3     | 2     | 4     | 1     | 2     | 1     | 2     | 1     | 6      | 2,44    |
| FPS ?      | A | 2     | 0     | 1     | 0     | 3     | 1     | 3     | 5     | 2      | 1,89    |
|            | V | 4     | 7     | 8     | 0     | 3     | 7     | 3     | 5     | 2      | 4,33    |
|            | T | 2     | 2     | 1     | 0     | 3     | 1     | 3     | 5     | 2      | 2,11    |
| Préférence |   | A,V,T | A,V,T | A,T,V | V,T,A | V,A,T | V,T,A | V,A,T | V,A,T | V,A,T  | V,A,T   |

**Tableau 1.** Le questionnaire final ; Les questions portent, dans l'ordre, sur le confort du dispositif, sur la gêne engendrée par le signal (SPF?) et sur la gêne engendrée par le film (FPS?) ; les différentes modalités sont la modalité Visuelle (V), Auditive (A) et Tactile (T) ; Les scores proposés allaient de 0 à 10, 10 représentant le pire.

### 4. Conclusion, discussion

L'objet de ce papier était de montrer la faisabilité, l'efficacité et l'acceptabilité de l'utilisation de trois modalités différentes d'un dispositif de réduction de surpressions durant une tâche attentive.

Tout d'abord, les résultats exposés montrent une cohérence globale (inter-modalités et inter-sujets) sur la perception et la compréhension du message de biofeedback. On notera le seul « faux-pas » du sujet paraplégique lors de la session tactile pendant laquelle celui-ci ne perçoit « que » 75% des signaux. Ce relatif mauvais score est à minorer par le fait qu'un signal non ou mal perçu pourra être recouvré au signal suivant.

Les résultats montrent aussi une cohérence intra-sujet entre les diverses modalités en terme d'efficacité. On peut donc en déduire que, peu importe la nature du signal, un tel dispositif peut être utilisé afin de réduire les surpressions emmagasinées jusqu'alors.

Enfin, en terme ergonomique, on notera qu'aucun dispositif n'est idéal aux yeux des sujets : le visuel et l'auditif, qui sont les modalités préférées des sujets, souffrent d'une interférence avec le visionnage du film. La modalité tactile, quant à elle, si elle est considérée comme celle interférant le moins avec le film, souffre d'un confort jugé très mauvais. Ceci est certainement dû au fait que, malgré les efforts de notre laboratoire dans l'ergonomie du TDU (notamment en rendant celui-ci entièrement sans fil), ce prototype reste très intrusif.

C'est pourquoi notre groupe envisage de tester de nouveaux dispositifs informatifs, ceux-ci pouvant être vibro-tactiles, voir mutlimodaux : par exemple une montre ou un appareil de type PDA ou iPhone équipé d'un mini-moteur vibrant pourrait alerter (par vibration) le sujet qu'un message urgent est à consulter. Le message en lui-même pourrait apparaître sur la montre ou le PDA sous forme textuel ou iconographique. Ceci est la proche perspective envisagée suite aux résultats présentés dans cet article.

## 5. Bibliographie

Bach-y-Rita P., Collins C.C., Saunders F., White B., Scadden L., « Vision substitution by tactile image projection..», *Nature*, 1969, vol 221, p. 963-964.

Bach-y-Rita P., Kaczmarek K.A., Tyler M.E., Garcia-Lara J., « Form perception with a 49-point electrotactile stimulus array on the tongue: A technical note.», *Journal of Rehabilitation Research and Development*, 1998, vol 35, p.427-430.

Barrois B., « Epidémiologie et coût de l'escarre : Epidémiologie des escarres, épidémiologie des escarres, méthodologie et résultants dans les pays occidentaux.», dans Colin D, Barrois B, Pelissier J. *L'escarre*. Paris: Masson, 1998.

Byrne D.W., Salzberg C.A., « Major risk factors for pressure ulcers in the spinal cord disabled: a literature review.», *Spinal Cord,* 1996, vol 34 n°5, p. 255-263.

Chen Y., DeVivo M.J., Jackson A.B., « Pressure Ulcer Prevalence in People With Spinal Cord Injury: Age-Period-Duration Effects.», *Archive of Physical Medecine and Rehabilitation, 2005*, vol 86 n°6, p. 1208-1213.

Garber S.L., Rintala D.H., « Pressure ulcers in veterans with spinal cord injury: a retrospective study. », *J Rehabil Res Dev.*, vol. 40, n° 5, 2003, p. 433-441.

Reddy M., Gill S.S., Rochon P.A., « Preventing Pressure Ulcers: A Systematic Review.», *JAMA,* 2006, vol 296 n°8, p. 23-30.

Salzberg C.A., Byrne D.W., Cayten C.G., van Niewerburgh P., Murphy J.G., Viehbeck M., « A new pressure ulcer risk assessment scale for individuals with spinal cord injury.. », *J Phys Med Rehabil.,* 2003, vol. 75, n° 2, p. 96-104.